\documentclass[
 reprint,
 nofootinbib,
 amsmath,amssymb,
 aps,
]{revtex4-1}

\pdfoutput=1

\usepackage{atlasphysics}
\usepackage{graphicx}
\usepackage{amssymb}
\usepackage{lineno}
\usepackage{comment}
\usepackage{multirow}
\usepackage{subfigure}
\usepackage{dcolumn}
\usepackage[hyperindex,breaklinks]{hyperref}

\newcommand{\tpion}{\ensuremath{\pi_{\mathrm{T}}}}

\newcommand{\trho}{\ensuremath{\rho_{\mathrm{T}}}}
\newcommand{\tomega}{\ensuremath{\omega_{\mathrm{T}}}}
\newcommand{\techa}{\ensuremath{a_{\mathrm{T}}}}
\newcommand{\Zgamma}{\ensuremath{\Zboson\gamma}}
\newcommand{\Wgamma}{\ensuremath{\Wboson\gamma}}

\newcommand{\ptp}{\ensuremath{p_{\mathrm{T}}}}

\usepackage{preprintcover}
\PreprintCoverPaperTitle{Search for new resonances in $W\gamma$ and $Z\gamma$ Final States in $pp$ Collisions at $\sqrt{s}=8$~\TeV{} with the ATLAS Detector}
\PreprintIdNumber{CERN-PH-EP-2014-159}
\PreprintCoverAbstract{
This Letter presents a search for new resonances decaying to final states with a vector boson produced in association 
with a high transverse momentum photon, $V\gamma$,
 with $V= W(\rightarrow \ell \nu)$ or $Z(\rightarrow \ell^{+} \ell^{-})$, where $\ell = e$ or $\mu$. The measurements
use $20.3$~fb$^{-1}$ of proton--proton
collision data at a center--of--mass energy of $\sqrt{s}=8$~\TeV{} recorded with the ATLAS detector. 
No deviations from the Standard Model expectations are found, 
and production cross section limits are set at 95\% confidence level. Masses of the hypothetical $\techa$ and $\tomega$ states of a benchmark
	Low Scale Technicolor model are excluded in
the ranges $[275,960]\,\GeV$ and $[200,700]\cup[750,890]\,\GeV$, respectively. Limits at 95\% confidence level on the production cross section of a singlet scalar resonance decaying to $Z\gamma$ final states have also been
obtained for masses below $1180\GeV$.
}
\PreprintJournalName{PLB}

\begin{document}

\title{Search for new resonances in $W\gamma$ and $Z\gamma$ Final States in $pp$ Collisions at $\sqrt{s}=8$~\TeV{} with the ATLAS Detector}

\author{The ATLAS Collaboration}
\address{}

\begin{abstract}
This Letter presents a search for new resonances decaying to final states with a vector boson produced in association 
with a high transverse momentum photon, $V\gamma$,
 with $V= W(\rightarrow \ell \nu)$ or $Z(\rightarrow \ell^{+} \ell^{-})$, where $\ell = e$ or $\mu$. The measurements
use $20.3$~fb$^{-1}$ of proton--proton
collision data at a center--of--mass energy of $\sqrt{s}=8$~\TeV{} recorded with the ATLAS detector. 
No deviations from the Standard Model expectations are found, 
and production cross section limits are set at 95\% confidence level. Masses of the hypothetical $\techa$ and $\tomega$ states of a benchmark
	Low Scale Technicolor model are excluded in
the ranges $[275,960]\,\GeV$ and $[200,700]\cup[750,890]\,\GeV$, respectively. Limits at 95\% confidence level on the production cross section of a singlet scalar resonance decaying to $Z\gamma$ final states have also been
obtained for masses below $1180\GeV$.
\end{abstract}

\maketitle
\section{Introduction}
\label{sec:intro}
The search for diboson resonances is an essential step in exploring the source of Electroweak Symmetry Breaking (EWSB). 
The observation of a Higgs boson decaying to $\gamma\gamma$, $ZZ$ and $W^+W^-$ final states, reported by the ATLAS and CMS Collaborations~\cite{ATLAS_Higgs,CMS_Higgs}, represents a milestone in particle physics history. However, the precise nature of the observed Higgs boson is not well known, and a dynamical mechanism of EWSB and fermion mass generation may yet involve 
a variety of heavy bosons of spin-0 or spin-1. 

This Letter presents a search for narrow scalar and vector heavy resonances decaying to $W\gamma$ or $Z\gamma$ final states, where the $W$ and $Z$ bosons decay to leptons ($e$ or $\mu$). 
The existence of a new vector or scalar resonance coupling to a boson pair $W\gamma$ or $Z\gamma$ is predicted by many physics scenarios, including various new physics models with scalar particles~\cite{scalarKK,scalarKK2} and vector particles~\cite{LSTC}.

The Low Scale Technicolor (LSTC)~\cite{LSTC} model is used as a benchmark model for the search for spin-1 resonances decaying to $W\gamma$ and $Z\gamma$ final states explored in this paper. The discovery of a Higgs boson, with its parameters in agreement with the Standard Model (SM) predictions, does not exclude the full phase space of the LSTC model, and the basic phenomenology would remain valid for a Technicolor model with a light composite Higgs boson.
To minimize the model dependence of the search results,
the signal cross section is measured within a well defined fiducial region.
The predicted new bound states of the lightest doublet of technifermions, the technimesons $\techa,\tomega, \trho$ and $\tpion$, generate a new phenomenology.
The mass splittings between the technimesons are set to be as follows: $m_{\rho_{\rm{T}}}=m_{\tomega}$, $m_{\techa}\approx{}1.1\times{}m_{\trho}$ and $m_{\rho_{\rm{T}}}-m_{\tpion}=m_{W}$~\cite{LSTC2}. The decays of technimesons to technipions are therefore kinematically forbidden. The technimesons mostly decay to pairs of electroweak gauge bosons, the most abundant decay channels being $\tomega \to
\Zgamma$, $\techa \to \Wgamma$ and $\trho \to \Wboson\Zboson, \Wgamma$. 
These technimeson resonances are expected to be narrow, with typical values $\Gamma(\trho, \tomega, \techa) \approx 1 \gev$.

A phenomenological model describing a singlet scalar particle $\phi$~\cite{benchmark_spin0a,benchmark_spin0b} is chosen as another benchmark in the search for spin-0 resonances decaying to $\Zgamma$. 
The neutral scalar could be composite, produced by a hypothetical new strong interaction.
It could be the pseudo--Goldstone boson playing an important role in the dynamical EWSB. This low energy effective model is independent of the underlying dynamical details and its Lagrangian can be written as follows:

\begin{multline}
\label{complag}
\mathcal{L}_{eff}= c_{g} \frac{4\pi\alpha_{s}}{\Lambda}\phi G^{a}_{\mu\nu}G^{a \mu\nu} + c_{W}\frac{4\pi\alpha_{em}}{\Lambda\cdot \sin^{2} \theta_{W}}\phi W^{a}_{\mu\nu}W^{a \mu\nu} \\
+ c_{B}\frac{4\pi\alpha_{em}}{\Lambda\cdot \cos^{2} \theta_{W}}\phi B_{\mu\nu}B^{\mu\nu}.
\end{multline}

Here $\alpha_{s}$ and $\alpha_{em}$ are, respectively, the couplings of the strong and electromagnetic interactions, $\Lambda$ is the cutoff scale and $\theta_{W}$ is the Weinberg angle. Moreover, $c_{g}$, $c_{W}$ and $c_{B}$ are the coupling coefficients between the scalar field $\phi$ and the gluon field strength $G^{a}_{\mu\nu}$, the $SU(2)$ field strength $W^{a}_{\mu\nu}$ and the $U(1)$ field strength $B_{\mu\nu}$ respectively. The scalar field $\phi$ interacts with the gauge boson pairs directly via dimension--$5$ operators, rather than via loop-induced effective interactions as in the case of the SM Higgs particle. Therefore, the gluon--gluon fusion production of $\phi$ and the branching ratio to $Z\gamma$ decay are significantly enhanced compared to the SM equivalent process $gg\to H \to Z\gamma$. No SM Yukawa interaction of $\phi$ with fermions is allowed so that there is no decay to $b\bar{b}$ or $t\bar{t}$ final states. The $WW$ and $ZZ$ decays of $\phi$ are largely suppressed due to its electroweak--singlet nature, compared with the SM Higgs boson. 

Previous limits on new resonances decaying to $W\gamma$ and $Z\gamma$ final states from $p\bar{p}$ and $pp$ production have been obtained at the Tevatron by the D$\O$ Collaboration~\cite{D0Paper} and at the Large Hadron Collider (LHC) by the ATLAS Collaboration~\cite{ATLASPaper}. At ATLAS, the production of $\techa{}$ and $\tomega{}$ for masses below 703~\GeV{} and 494~\GeV{}, respectively, was excluded within the LSTC benchmark parameters. The most stringent limits on LSTC have been set by the CMS~\cite{CMSPaper} Collaboration, excluding the production of $\trho$ for masses below $m_{\trho}<1.14~\TeV$, but using a slightly different choice of parameters.

\section{ATLAS detector and data sample}
\label{sec:detector}

The ATLAS detector~\cite{ATLASdetector} is composed of
an inner tracker detector (ID) surrounded by a thin
superconducting solenoid providing a 2~T axial magnetic field,
electromagnetic (EM) and hadronic calorimeters (HC), and a muon spectrometer (MS)
immersed in the magnetic field produced by a system of superconducting
toroids.  
The ID consists of three subsystems: the pixel and silicon
microstrip detectors cover the pseudorapidity\footnote{ATLAS uses a right-handed coordinate system with its origin at the nominal interaction point (IP) in the center of the detector and the $z$-axis along the beam pipe. The $x$-axis points from the IP to the center of the LHC ring, and the $y$-axis points upward. Cylindrical coordinates $(r,\phi)$ are used in the transverse $(x,y)$ plane, $\phi$ being the azimuthal angle around the beam pipe. The pseudorapidity is defined in terms of the polar angle $\theta$ as $\eta=-\ln\tan(\theta/2)$. The distance $\Delta R$ in the $\eta-\phi$ space is defined as $\Delta R = \sqrt{({\Delta \eta})^2 + ({\Delta \phi})^2}$.} range $|\eta|<2.5$, while the transition radiation tracker, made of straw tubes, has an acceptance range of $|\eta|<2.0$.
The calorimeter system covers the range $|\eta|<4.9$.
The highly segmented electromagnetic calorimeter plays a crucial role in electron
and photon identification. It is composed of lead absorbers with Liquid Argon (LAr) as the active
material and spans $|\eta|<3.2$.
In the region $|\eta|<1.8$, a pre-sampler detector using a thin layer of LAr is used to
correct for the energy lost by electrons and photons upstream of the calorimeter.
The hadronic tile calorimeter ($|\eta|<1.7$) is a steel/scintillating-tile detector and is located directly
outside the envelope of the barrel electromagnetic calorimeter.
The end--caps and forward calorimeters use LAr as the active material, with copper (EM) and tungsten (HC) as absorber materials.
The MS is composed of three large superconducting air--core toroid magnets of
eight coils each,
a system of three stations of high precision tracking chambers in the range $|\eta|<2.7$,
and a muon trigger system which covers $|\eta|< 2.4$.
The precision measurement
is ensured by monitored drift tubes and, 
at large pseudorapidities ($|\eta| > 2$), for the innermost layer, by cathode strip chambers.
The muon trigger system is composed of resistive plate chambers in the barrel region ($|\eta| < 1.05$),
and thin gap chambers in the end--cap ($1.05 < |\eta| < 2.4$).

The ATLAS trigger system has three distinct levels, L1, L2 and the event filter,
where each trigger level refines the decisions made at the previous level.
The data used for the present analysis were collected in 2012 from
$pp$ collisions at a center-of-mass energy of $8\TeV$ at the LHC.
The total integrated luminosity is 20.3~fb$^{-1}$ with an
uncertainty of 2.8$\%$~\cite{ATLASlumi2013}.
Events are selected by triggers requiring at least one identified
electron or muon.
The transverse energy ($E_{\mathrm{T}}$) threshold for the
single-lepton trigger is $24\GeV$.  The lepton trigger efficiencies 
are measured using $Z$ boson candidates
as a function of the transverse momentum $\ptp$ and $\eta$. 
The trigger efficiencies for the leptons
are approximately 70\% for muons with  $|\eta| <$ 1.05, 90\% for muons in the range 1.05 $< |\eta| <$ 2.4~\cite{ATLASmuontrigger}, and 95\% for electrons in the range  $|\eta|<$ 2.4.

\section{Signal and background simulated samples}
\label{sec:simulation}
 Monte Carlo (MC) event samples, including a full simulation~\cite{ATLASsimulation}
of the ATLAS detector with {\sc Geant4}~\cite{geant},
are used to compare the data to the signal and background expectations.
All MC samples are simulated with additional $pp$ interactions (pile-up) in the same and neighboring bunch crossings. 

The production and decay of neutral (\tomega{} $\to$ $Z\gamma$) and charged technimesons (\techa{} $\to$ $W\gamma$) in the LSTC model are handled by the {\sc Pythia6.426} generator~\cite{pythia}. The following parameters are used in the event generation: number of technicolors $N_{TC}=4$; techniquark charges $Q_{U}=1$ and $Q_{D}=0$ for the $Z\gamma$ final state and $Q_{U}=1/2$ and $Q_{D}=-1/2$ for the $W\gamma$ final state. With this parameterization of the techniquark charges, the dominant contribution to the $W\gamma$ final state is from \techa{} decay.
 By removing the \trho{} contribution, the model dependence that could result from having two nearby peaks in the benchmark signal is further reduced. The sine of the mixing angle between the technipions and the electroweak gauge boson longitudinal component is set to 1/3. As stated in the introduction, the mass splittings between the technimesons are set to: $m_{\rho_{\rm{T}}}=m_{\tomega}=m_{\techa}/1.1$, and $m_{\rho_{\rm{T}}}-m_{\tpion}=m_{W}$.

Simulations of the signals for singlet scalar particles are generated using {\sc Madgraph5}~\cite{Alwall:2007st} interfaced to {\sc Pythia8}~\cite{pythia8} for parton shower and fragmentation processes. The generation uses the leading-order (LO) parton distribution function (PDF) set {\sc MSTW2008LO}~\cite{mstw2008lo}.
Since the kinematic distribution of the $Z\gamma$ decay of a scalar boson is determined by its spin and CP properties in the narrow--width approximation, the gluon-gluon fusion production and the decay are simulated by using the SM Higgs process in {\sc Madgraph5}, fixing the resonance width to $5.75~\MeV$. 

The cross section of the singlet scalar process $gg \to \phi \to Z\gamma$ is calculated by choosing two sets of coupling parameters in Eq.~\ref{complag}.
In the parameter set (a),
the coupling coefficients are set to $c_{g}=c_{W}=-c_{B}=1/4\pi$, as suggested by Naive Dimensional Analysis (NDA)~\cite{NDAold}. 
Here, the relative phases are chosen to enhance the $\Zgamma$ coupling.
If a large new color number $N'_{c}=4$ of the underlying strong dynamics is considered in NDA~\cite{NDAnew, NDAnew2}, and larger couplings with the electroweak gauge bosons are assumed~\cite{benchmark_spin0b},
then larger values of the coefficients can also be possible, depending on the underlying theory. Thus another parameter set (b) is selected as follows: $c_{g}=1/2\pi, c_{W}=-c_{B}=1/\pi$.
In both parameterizations, the $Z\gamma$ decay rate is dominant over the $\gamma\gamma$ decay rate. The underlying dynamics scale is set to be $\Lambda=6~\TeV$, motivated by the Higgs boson couplings measurements~\cite{lambda6TeVa,lambda6TeVb}. The width of the singlet scalar for each of set (a) and (b) is well below the experimental resolution over the full mass range studied. 

The main background processes, SM $pp \to \ell^+ \ell^- \gamma$ and $pp \to \ell \nu  \gamma$ production, are modeled using the {\sc Sherpa}~(1.4.1) generator~\cite{sherpa1}. An invariant mass cut of $m(\ell^{+}\ell^{-})>40\GeV$ is applied at the generator level when simulating the $pp \to \ell^+ \ell^- \gamma$ process.
The {\sc CT10}~\cite{Lai:2010vv} PDF is used for samples generated with {\sc Sherpa}.
In the $W\gamma$ analysis, the $Z(\ell^{+}\ell^{-})$ and $Z(\tau^{+} \tau^{-})$ backgrounds are
modelled with {\sc Pythia8}. 
The final state radiation of photons from charged leptons is treated in {\sc Pythia8}
using {\sc Photos}.
{\sc Tauola}~(1.20)~\cite{tau} is used to model $\tau$ lepton decays. Events with one or more hard photons ($p_T>$10 \GeV{}) in $Z(\ell^{+}\ell^{-})$ and $Z(\tau^{+} \tau^{-})$ MC simulations are removed, in order to avoid overlaps with $Z\gamma$ MC samples. 
In the $Z\gamma$ analysis, $Z(\ell^{+}\ell^{-})$ and $Z(\tau^{+} \tau^{-})$ MC samples are used to cross check the data-driven $Z+$jet background estimation.

The {\sc Powheg}~\cite{powheg} generator is used to simulate $t\bar{t}$ production, and is
interfaced to {\sc Pythia8} for parton showering and fragmentation. 
The single top quark, $WW$, $WZ$ and $ZZ$ processes are modelled by
{\sc MC@NLO}~(4.02)~\cite{MCatNLO1,MCatNLO2}, interfaced to {\sc Herwig}~(6.520)\cite{herwig}
for parton showering and fragmentation processes and to {\sc Jimmy}~(4.30)~\cite{jimmy} for underlying event simulation.
The LO {\sc MRST2007}~\cite{mrst2007} PDF set is used to simulate the
$Z(\ell^{+}\ell^{-})$ backgrounds, and the {\sc CT10} PDF set is used in simulating $t\bar{t}$, single top quark, $WW$, $WZ$ and $ZZ$ production.

\section{Physics object reconstruction and selection} 
\label{sec:obj}
The $W$ and $Z$ bosons are reconstructed from their
leptonic decays and are required in addition to an isolated high $E_{\mathrm{T}}$ photon.
Furthermore, collision events are selected by requiring a reconstructed vertex with at 
least three charged particle tracks with $p_{\mathrm{T}}>0.4\GeV$.
In events with more than one candidate vertex, the one with the highest sum of the $p_{\mathrm{T}}^{2}$ of associated tracks is chosen.

An electron candidate is obtained from an energy cluster in the EM calorimeter
associated with a reconstructed track in the ID.
The transverse energy of electrons is required to be greater than 25~\GeV{}.
The electron cluster must lie 
within the overall fiducial acceptance of the EM calorimeters and the ID. In the
$W\gamma$ final states, the electrons found in the transition region between the barrel and end-cap EM calorimeters are 
removed in order to improve the $E_{\mathrm{T}}^{\mathrm{miss}}$ resolution.
The acceptance requirements are $|\eta|<1.37$ or $1.52<|\eta|<2.47$ 
for $W\gamma$ events and $|\eta| < 2.47$ for $Z\gamma$ events.
At the electron track's closest approach to the primary vertex, the ratio of the transverse impact parameter
$d_{0}$ to its uncertainty (the $d_{0}$ significance) must be smaller than 6.5,
and the longitudinal impact parameter $|z_{0}|$ must be less than 10~mm.
Tight
electron identification, as defined in Ref.~\cite{atlas_electron}, is used in the
$W(e \nu)\gamma$ analysis,
whereas loose identification~\cite{atlas_electron} is used to select electrons in the
$Z(e^+ e^-)\gamma$ analysis.
In order to reduce the background due to a jet misidentified as an electron, a calorimeter--based isolation requirement is applied to the electron candidate. The normalized calorimetric isolation ($R_{\mathrm{calo}}^{\mathrm{iso}}$) for electrons must satisfy $R_{\mathrm{calo}}^{\mathrm{iso}}<0.20$. It is computed as the sum of the positive-energy topological clusters with a reconstructed barycenter falling in a cone of $\Delta R < 0.2$  around the candidate electron cluster divided by the electron $E_{\mathrm{T}}$. The energy deposited by the electron is subtracted.

In both $W\gamma$ and $Z\gamma$ analyses, muon candidates are identified by associating complete tracks or track segments in the
MS to tracks in the ID~\cite{WZpaper}.
Each of these muon candidates corresponds to a combined ID and MS track originating from the primary vertex
with transverse momentum $p_{\mathrm{T}}>25\GeV$ and $|\eta|<2.4$. 
In the $Z\gamma$ analysis, muon candidates can also be formed from tracks reconstructed either in the ID or in the MS.
This allows one to increase the signal efficiency by extending the muon $|\eta|$ acceptance up to $2.7$. 
In the central barrel region $|\eta|<0.1$, 
which lacks MS coverage, ID tracks are identified as muons based on the associated energy deposits in the calorimeter. These muons are required to have a calorimeter energy deposit consistent with that of a minimum ionizing particle.
Muons are required to be isolated by imposing $R_{\mathrm{track}}^{\mathrm{iso}}<0.3$ ($0.15$ for muons without an ID track),
where $R_{\mathrm{track}}^{\mathrm{iso}}$ is the sum of the $p_{\mathrm{T}}$ of the tracks
in a $\Delta R = 0.2$ cone around the muon direction, excluding the track of the muon,
divided by the muon $p_{\mathrm{T}}$.
The $d_{0}$ significance must be smaller than 3.5, and $|z_{0}|$ must be less than 10~mm.

Overlap removal is applied to electrons and muons that satisfy all selection criteria and share the same ID track, in order to avoid double counting.  In particular, if  the muon is identified by the MS, then the electron candidate is discarded, otherwise the muon candidate is rejected. 

Photon candidate reconstruction relies on clustered energy deposits in the EM calorimeter with $E_{\mathrm{T}}> 40$~\GeV{} in the
range $|\eta|<2.37$ (excluding the calorimeter transition region $1.37<|\eta|<1.52$).
Clusters without matching tracks are directly classified as unconverted photon candidates.
Clusters that are matched to tracks which originate from reconstructed conversion vertices in the ID
or to tracks consistent with coming from a conversion are considered as converted photon candidates. 
Both unconverted and converted candidates are used in the present analysis.
Tight requirements on the shower shapes~\cite{atlas_electron} are
applied to suppress the background from multiple showers produced in
meson (e.g. $\pi^{0}$, $\eta$) decays. To further reduce this
background, a photon isolation requirement $E_{\mathrm{T}}^{\mathrm{iso}}<4$~\GeV{} is applied.
The isolation variable $E_{\mathrm{T}}^{\mathrm{iso}}$ is the total transverse energy recorded in the
calorimeters within a cone of $\Delta R = 0.4$ around the photon position
excluding the energy deposit of the photon itself.
The value of $E_{\mathrm{T}}^{\mathrm{iso}}$ is corrected for leakage from the photon energy cluster core into the isolation cone and for contributions 
from the underlying event and pile-up~\cite{atlas_electron,ATLASPhotonIDConf}.

Jets are reconstructed from calibrated topological clusters built from energy deposits~\cite{ATLASdetector} in the calorimeter using the anti-$k_t$ jet clustering algorithm~\cite{antikt} with radius parameter $R=0.4$.
The selected jets are required to have $p_{\mathrm{T}}>25$~\GeV{} with $|\eta| < 4.5$,
and to be well separated from the lepton and photon candidates
($\Delta R(e/\mu/\gamma,$ jet)$>0.3$).

The $W\gamma$ final state has missing transverse momentum, $E_{\mathrm{T}}^{\mathrm{miss}}$~\cite{METpaper},
due to the undetected neutrino from the $W$ boson decay.
Its magnitude and direction are measured from the vector sum of
the transverse momentum vectors associated with clusters of energy reconstructed
in the calorimeters with $|\eta|<4.9$.
A correction is applied to the energy of those clusters that are associated with
a reconstructed physical object (jet, electron, $\tau$-lepton, photon).
Reconstructed muons are also included in the sum, and any calorimeter energy deposits
associated with them are excluded to avoid double counting.

\section{$W\gamma$ and $Z\gamma$ event selection}
\label{sec:eventselection}

The $\ell \nu \gamma$ candidate events are selected by requiring exactly one lepton with $p_{\mathrm{T}}>25$~\GeV{}, at least one photon with $E^{\gamma}_{\mathrm{T}}>40$~\GeV{} and $E_{\mathrm{T}}^{\mathrm{miss}}$ above 35~\GeV{}.  
In addition, the transverse mass \footnote{The transverse mass is defined as $m_{\mathrm{T}}= \sqrt{2p_{\mathrm{T}}(\ell) \cdot E_{\mathrm{T}}^{\mathrm{miss}} \cdot (1-\cos{\Delta{\phi}})}$, 
where $\Delta{\phi}$ is the azimuthal separation between the directions of the lepton
and the missing transverse momentum vector.} of the lepton--$E_{\mathrm{T}}^{\mathrm{miss}}$ system 
 is required to be greater than 40~\GeV{}.
A $Z$-veto requirement is applied in the electron channel of the $W\gamma$ analysis by requiring
that the electron--photon invariant mass be not within 15~\GeV{} of the $Z$ boson mass.
This is used to suppress the background where one of the electrons from the $Z$ boson decay
is mis-identified as a photon.

The $\ell^+ \ell^- \gamma$ candidates are selected by requiring two oppositely charged same--flavor leptons with an invariant mass between 65 and 115~\GeV{}, and at least one photon with $E^{\gamma}_{\mathrm{T}}> 40$~\GeV{}. For $\mu^{+}\mu^{-}\gamma$, at least one of the two muons must be reconstructed both in the ID and the MS. If more than one $Z$ candidate is found in the same event, the one with mass closest to the $Z$ mass is retained.

In both the $W\gamma$ and $Z\gamma$ analyses, a selection requirement $\Delta R(\ell,\gamma)>0.7$ is applied to suppress the contributions from photons radiated from the selected leptons. If more than one photon passes all of the selection criteria, the one with the highest $\ptp$ is selected.

\section{Analysis Strategy}
\label{sec:strategy}

A search for a signal is performed with an unbinned maximum likelihood fit of the mass spectra of the $\ell\nu\gamma$ and $\ell^{+}\ell^-\gamma$ systems. 
In order to select and validate the probability density function used to model the background shape, a study of the background composition is also performed.
In the final results, the background normalization and shape are extracted by performing a signal-plus-background maximum likelihood fit directly to the data spectrum.

\section{Study of the background composition}
\label{sec:background}

The composition of the backgrounds affecting the $\ell \nu \gamma$ and $\ell^{+} \ell^{-} \gamma$ searches are studied by using a combination of simulation and data--driven methods.
The dominant contribution is the SM $V\gamma$ production, then come other processes, like $V+$jets, $\gamma$+jets, $t\bar{t}$ and other diboson final states, such as $WW$ and $WZ$.
The SM $V\gamma$ backgrounds are estimated using {\sc Sherpa} MC samples. A control region is defined to validate the MC modeling. The events in the control regions are selected by imposing the nominal event selection criteria, except that the photon $E^{\gamma}_{\mathrm{T}}$ is required to be between 15 and 40~\GeV{}. The comparison of the relevant kinematic distributions, such as the transverse momenta of the leptons and photons and the transverse missing energy corresponding to the undetected neutrino, in the control regions shows that the $V\gamma$ background is well modeled by the MC simulation.
The backgrounds estimated from data are $Z/W+$jets and $\gamma+$jets
in the $\ell \nu \gamma$ final state, and $Z+$jets in the $\ell^{+} \ell^{-} \gamma$ final state.
The data--driven backgrounds are estimated by using a two--dimensional sideband method, as described in Ref.~\cite{WZgpaper}.
The remaining backgrounds, \emph{i.e.} $t\bar{t}$, single top, $WW$, $WZ$, and $ZZ$ are estimated with MC simulations.

\begin{figure*}[htbp]
  \centering
  \subfigure[]{\includegraphics[trim=35 20 60 5,clip=true,width=1.\columnwidth]{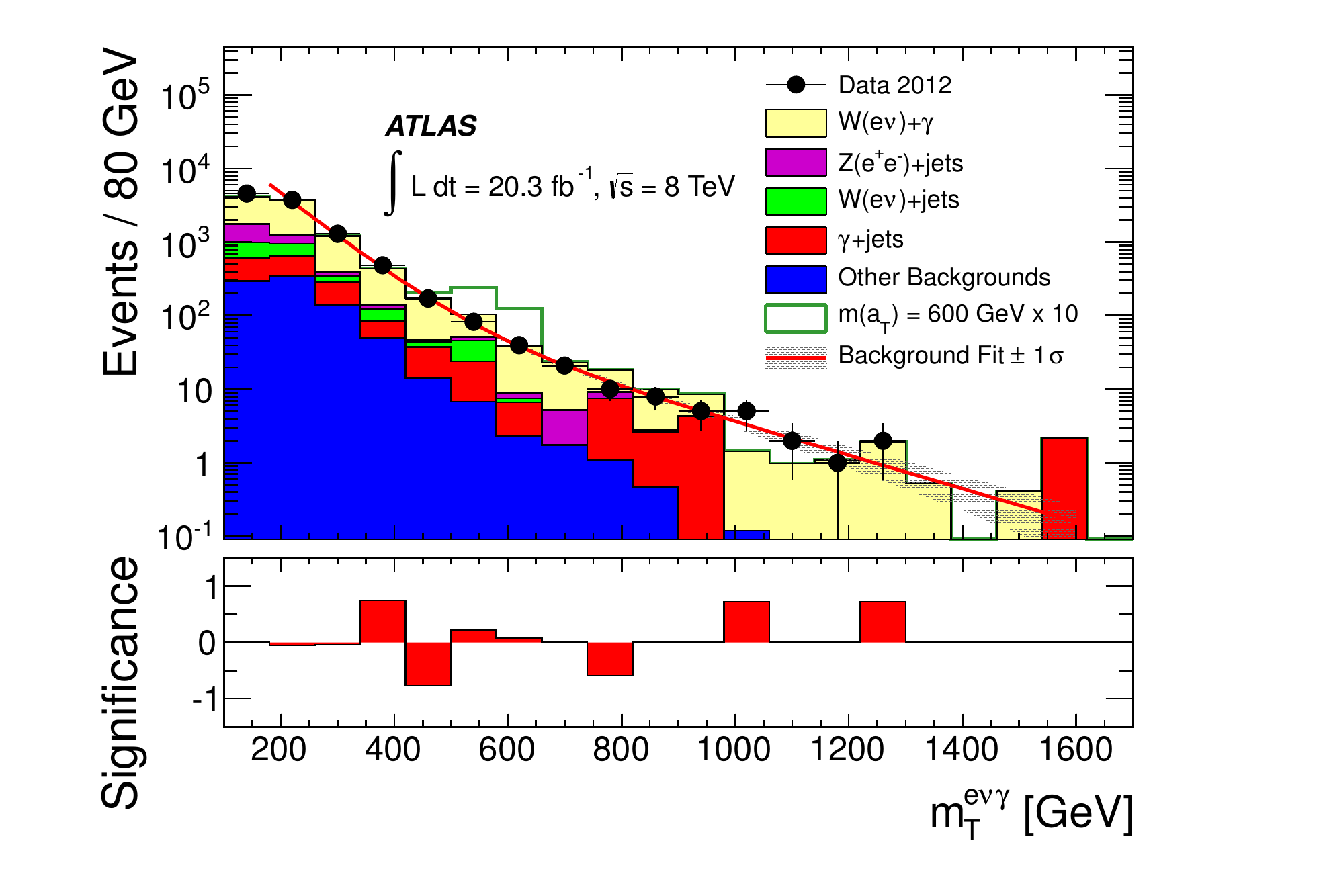}}
  \subfigure[]{\includegraphics[trim=35 20 60 5,clip=true,width=1.\columnwidth]{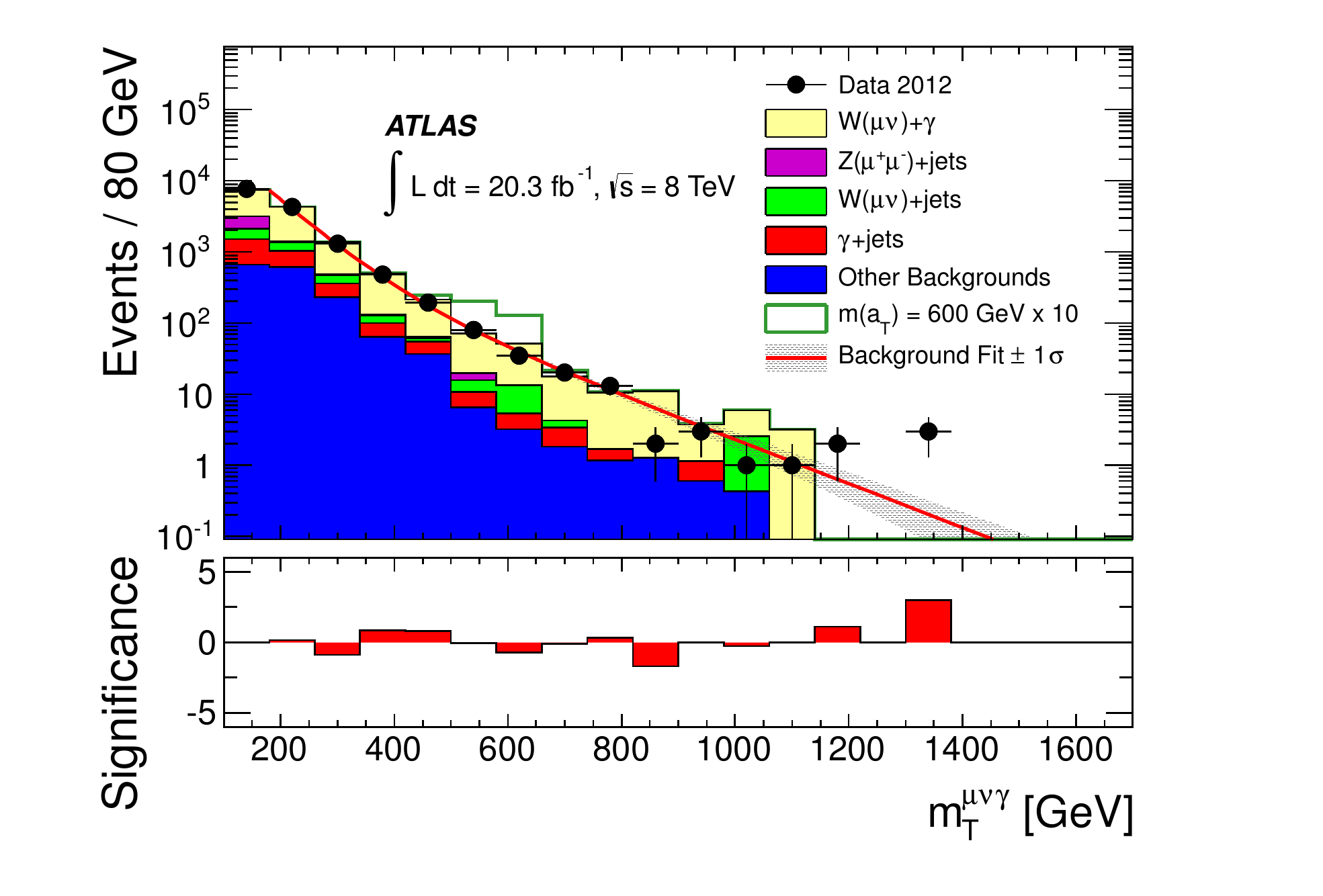}}
  \caption{Three-body transverse mass distribution for the $e\nu\gamma$ (a) and $\mu\nu\gamma$ (b) final states. The background-only fit to the data and the one standard deviation uncertainty band are shown. 
The significance (see text for details) as a function of $m_{\mathrm{T}}^{\ell\nu\gamma}$ in the lower panel  includes only statistical uncertainties. 
The expected signal for a resonance mass of $600~\GeV$ is superimposed.}
  \label{fig:AllMassDistributionsW}
\end{figure*}

\begin{figure*}[htbp]
  \centering
\subfigure[]{\includegraphics[trim=35 20 60 5,clip=true,width=1.\columnwidth]{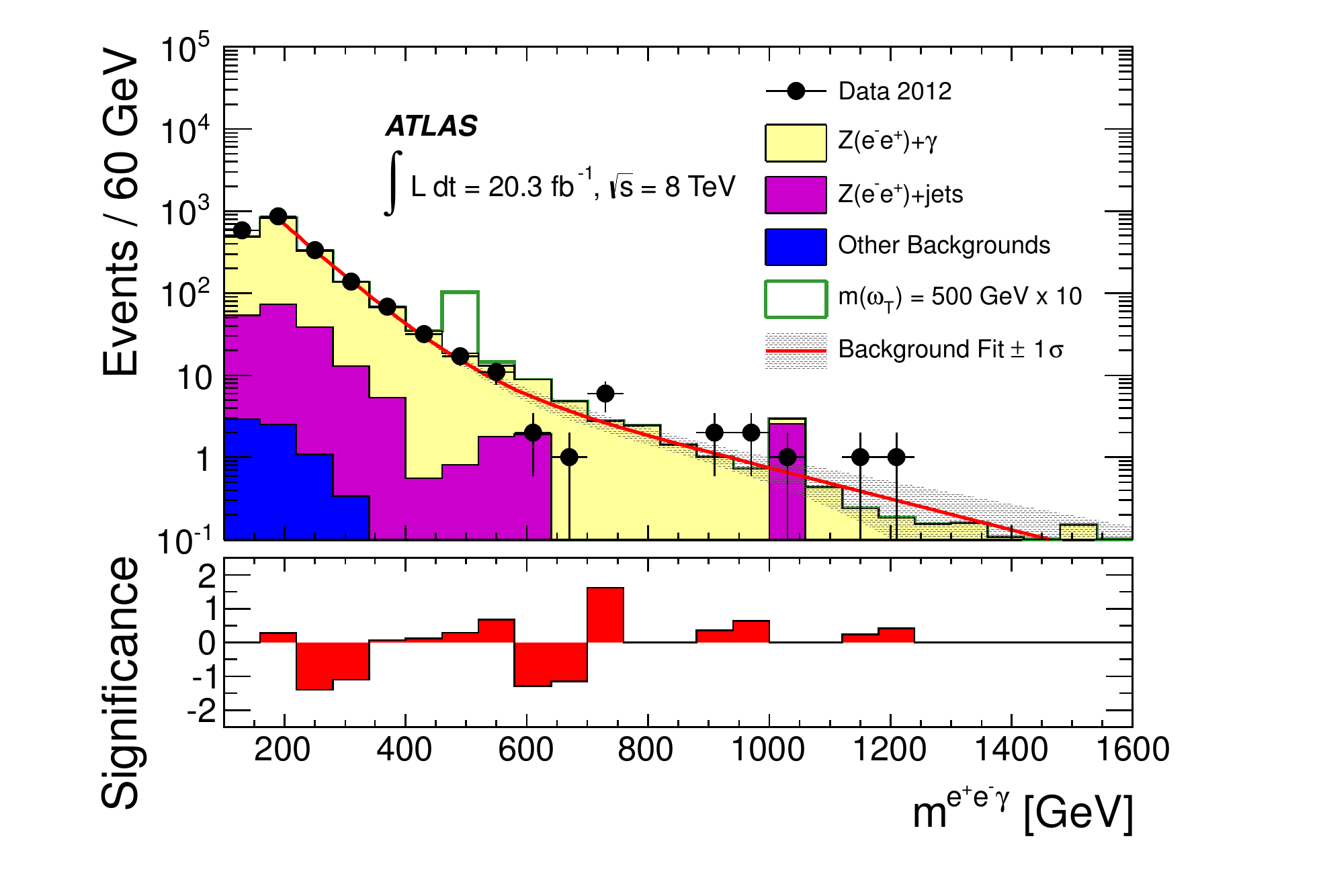}}
\subfigure[]{\includegraphics[trim=35 20 60 5,clip=true,width=1.\columnwidth]{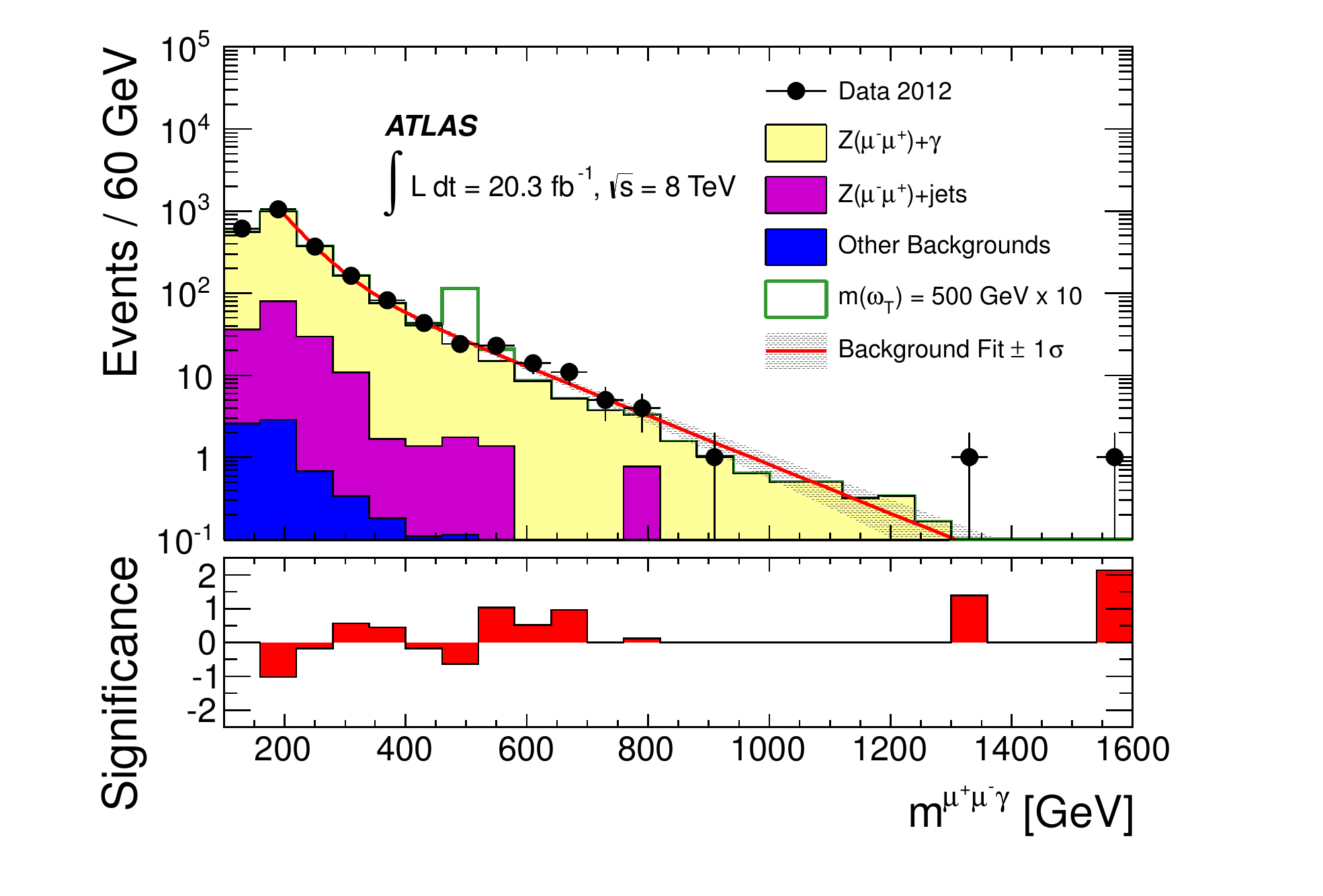}}
  \caption{Three-body invariant mass distribution for (a) the $e^{+}e^{-}\gamma$ and (b) $\mu^{+}\mu^{-}\gamma$ final states. The background-only fit to the data and the one standard deviation uncertainty band are is shown.
The significance (see text for details) as a function of $m^{\ell^+\ell^-\gamma}$ in the lower panel includes only statistical uncertainties. 
The expected signal for a resonance mass of $500~\GeV$ is superimposed.}
  \label{fig:AllMassDistributionsZ}
\end{figure*}

The background prediction obtained in all four final states is shown in Table~\ref{tab:bg}.
The distribution of the $W\gamma$ system three-body transverse mass \footnote{The $W\gamma$ system three-body transverse mass is defined as:
\begin{eqnarray}
 (m_{\mathrm{T}}^{ \ell \nu \gamma})^{2} &=& ( \sqrt{m^2_{\ell \gamma}+|\vec{p}_{\mathrm{T}}(\gamma)+\vec{p}_{\mathrm{T}}(\ell)|^2}+E_{\mathrm{T}}^{\mathrm{miss}})^2      \nonumber \\
    && -| \vec{p}_{\mathrm{T}}(\gamma)+\vec{p}_{\mathrm{T}}(\ell)+\vec{E}_{\mathrm{T}}^{\mathrm{miss}}|^2\nonumber  .\\ 
\label{equ:MT3}
\end{eqnarray}
}
for the selected $W\gamma$ events is shown in Fig.~\ref{fig:AllMassDistributionsW}, while Fig.~\ref{fig:AllMassDistributionsZ} presents the $Z\gamma$ invariant mass
distribution. Here, data are compared to the sum of the SM expected backgrounds and show fair agreement. 
The significance, shown as a function of the mass in the lower panels, only includes statistical uncertainties and is defined as the $z$--value between the data points and the fit expectation, which is the deviation at the right of the mean of a Gaussian distribution, expressed in units of standard deviations, as described in Ref.~\cite{zvalue}. 
The event rate in the high mass region has a significance below three standard deviations, and is attributed to a statistical fluctuation.

\begin{table*}[htbp]
 \begin{tabular}{ccccc}
       \hline
       Background & $ e\nu \gamma$  & $\mu\nu \gamma$ &$e^{+}e^{-}\gamma$ & $\mu^{+}\mu^{-}\gamma$ \\
       \hline
       $W(\ell \nu)\gamma$          & $6100 \pm 50  $       & $8710 \pm 150  $ & -- & --\\  
       $Z(\ell \ell)\gamma$         & \multirow{2}{*}{$1160 \pm 40$}& \multirow{2}{*}{$1100 \pm 40$} & $ 1770 \pm 20$ & $2170 \pm 30$\\  
       $Z(\ell \ell) +$jets       &         &    & $180 \pm 90 \pm 70 $ & $160 \pm 90 \pm 50 $  \\  
       $W(\ell \nu)$+jets           & $790 \pm 50\pm 330$ & $1110 \pm 70 \pm 310$ & -- & --\\  
       $\gamma$+jets                & $870\pm 60\pm 300$  & $1460 \pm 110\pm 510$ & -- & --  \\  
       $t\bar t$                    & $700 \pm 20$          & $1310 \pm 20$ & $4 \pm 1$         & $5 \pm 1 $  \\  
       other backgrounds            & $160  \pm 10$          & $310  \pm 11 $ & $2.4 \pm 0.3$ & $1.7 \pm 0.2$ \\ 
\hline
       Total expected               & $9780 \pm 80 \pm 450 $   & $14000 \pm 130 \pm 620 $ & $1960 \pm 90 \pm 80 $  & $2340 \pm 90 \pm 50 $                   \\
       \hline
       Total observed &    10437                  &   14082 & 2086    &  2429  \\ \hline

\hline
\end{tabular}
\caption{
Total number of events passing the $W/Z\gamma$ selection requirements in the data,
and expected number of SM background events in the $e\nu \gamma$, $\mu\nu \gamma$, $e^{+}e^{-}\gamma$ and $\mu^{+}\mu^{-}\gamma$ final states. The first uncertainty is statistical and, when present, the second uncertainty shows an estimate of the systematic effects. The systematic uncertainty is only presented for $W+$jets and $\gamma+$jets background in $W\gamma$ analysis, and for $Z+$jets background in $Z\gamma$ analysis, since the uncertainty on these background estimates dominates the uncertainty of the total background estimate. The ``other backgrounds'' include contributions from single top, $WW$ and $WZ$ production.}
  \label{tab:bg} 
\end{table*}

\section{Fit model and statistical methods}
\label{sec:fitstat}

The signal and background parameterizations have been chosen to best describe the mass distribution shapes of, respectively, fully simulated MC samples and a mixture of MC samples and data--driven estimates. 
The fit to data takes into account all systematic uncertainties, discussed in Sec.~\ref{sec:syst}.

\subsection{Background modeling}
\label{sec:bkg_model}
All the mass distributions, shown in Figs.~\ref{fig:AllMassDistributionsW} and~\ref{fig:AllMassDistributionsZ}, have a broad maximum in the low mass region due to the kinematic cuts used. The shape of the SM background is thus parameterized 
only for masses larger than 180~\GeV{}.

A probability density function is chosen to describe the expected SM background mass distribution for the range, 
between $180~\GeV$ and $1.6~\TeV$. 
The variation of the background composition, within systematic uncertainties, presented in Sec.~\ref{sec:background} is also taken into account in evaluating the background parametrization systematics.
For both the $W\gamma$ and $Z\gamma$ final states, the probability density function best describing the SM background distribution, in the chosen range, is found to be the sum of two exponentials. The impact of the choice of the background model on the final result has been studied, as discussed in Sec.~\ref{sec:syst}.

The results of the unbinned fit to the data can 
be seen as the solid curve on each of the mass spectra in Fig.~\ref{fig:AllMassDistributionsW} and~\ref{fig:AllMassDistributionsZ}. 
A good description of data is found, and the Pearson's $\chi^{2}$ per degree of freedom  obtained for the background-only fit is close to unity for all distributions.

\subsection{Signal modeling}
\label{sec:sig_model}
The parameterizations of the signal shapes have been obtained from the benchmark signal MC samples. For all $W/Z \gamma$ spin-1 and $Z\gamma$ spin-0 resonances considered, the signal mass distribution is described by the sum of a Crystal-Ball function (CB)~\cite{CrystalBall1,CrystalBall2,CrystalBall3}, 
which reproduces the core mass resolution plus a non-Gaussian tail for low mass values, and a small wider Gaussian component that takes into account outliers in the mass distribution. The mean value of the CB and the Gaussian are chosen to be the same.

For simulated events, the mean fitted mass is found to be within $2\GeV$ of the generated resonance mass for the $Z\gamma$ final state.  At the reconstruction level, the full width at half maximum of the $Z\gamma$ signal grows approximately linearly from $7\GeV$ 
to $38\GeV$ in the electron channel, and from $7\GeV$ to $100\GeV$ in the muon channel, for masses generated between $200\GeV$ and $1.6\TeV$. The full width at half maximum of the $W\gamma$ transverse mass also grows approximately linearly, and goes from $50\GeV$ 
to $100\GeV$ in the electron channel, and from $45\GeV$ to $190\GeV$ in the muon channel, for masses generated between $200\GeV$ and $1.6\TeV$. These widths are completely dominated by reconstruction effects, since we assume the hypothetical signals to be narrow.

\subsection{Statistical method}
\label{sec:statmeth}
The search is conducted by performing a scan of the $m_{Z\gamma}$ and $m^{\mathrm{T}}_{W\gamma}$ distributions using a probability density
function describing both the background shape and a signal peak at a given mass. The step size used in the scan is $20~\GeV$ 
for the $W\gamma$ channel and $10~\GeV$ for the $Z\gamma$ channel. The difference in the scan step takes into account the different resolutions of a hypothetical peak in the two channels. 
The scan starts at a signal mass of $275~\GeV$ for the $\Wgamma$ final state, and of $200~\GeV$ for the $\Zgamma$ final state, and goes up to $1600~\GeV$.
Each fit is performed in a mass range of $[180,1600]~\GeV$ for both $\Zgamma$ and $\Wgamma$ final states, in order to ensure that there are enough events in the sidebands of a hypothetical signal peak.

The background normalization and the two exponential coefficients of the SM background 
probability density function are free to vary in the fit.
The parameters of the signal probability density functions are fixed to their nominal values from fits to the MC signal model, except for the normalization of 
the signal and for the nuisance parameters that account for the systematic uncertainties on the signal event rate and resolution, 
described below. The systematic uncertainty related to the choice of parameterization for the background, on the other hand, has been estimated 
by testing different parameterizations and verifying that the extracted background and signal yields do not vary significantly, and was found 
to be negligible with respect to the other systematic and statistical uncertainties considered. 

The signal yield, $N_{S}$, depends on the parameter of interest of the fit, i.e. the signal fiducial
cross section $\sigma_{\mathrm{Fid}}$ at a given signal mass, as follows:
\begin{equation}\label{eq:fitproc}
 N_{S}=\sigma_{\mathrm{Fid}}\cdot\epsilon_{\mbox{\tiny{Reco}}}\cdot{}\int{}{\cal L}dt,
\end{equation}
where the factor $\epsilon_{\mbox{\tiny{Reco}}}$ is the signal reconstruction efficiency, 
defined as the number of reconstructed signal events passing the full event selection divided
by the number of events generated in the fiducial region defined in Table~\ref{tab:fiducialcut}. 
The signal reconstruction efficiency has been estimated by using signal MC simulated samples, and accounts for effects due to 
the detector resolution on the lepton and photon transverse momentum and energies, 
and on the missing transverse energy. 
The parameter $\epsilon_h^p$ in the table is defined at particle level as the sum of the energy carried by final state particles in a $\Delta R=0.4$ cone around the photon direction (not including the photon) divided by the energy carried by the photon. To account for the effect of final-state QED radiation, the energy of the generated lepton at particle level is defined as the energy of the lepton after radiation plus the energy of all radiated photons within a $\Delta R=0.1$ cone around the lepton direction.
The $\sigma_{\mathrm{Fid}}$ parameter is the same for both muon and electron channels in each final state,
and is thus determined by simultaneously performing the fit in 
the two channels, so that the results obtained are less sensitive to statistical fluctuations.\\

\begin{table}[!htbp]
  \centering 
  \begin{tabular}{lcccccc}
    \hline 
Cuts   &  $\ell \nu\gamma$  & $\ell^+ \ell^- \gamma$ \\
\hline
Lepton  & $p_{\mathrm{T}}^{\ell} >25\GeV{}$ & $p_{\mathrm{T}}^{\ell} >25\GeV{}$ \\
        & $|\eta_{\ell}|<2.47$  & $|\eta_{\ell}|<2.47$ \\
        & $N_\ell =1$& $N_{\ell^+} =1 , N_{\ell^-}=1$ \\
Neutrino   & $p_{\mathrm{T}}^\nu>35\ \GeV{}$    & ---  \\
\hline
Boson &     $m_{\mathrm{T}}^{W}>40$~\GeV{}   &   $65 < m^{Z}< 115~\GeV$  \\
\hline
Photon  & \multicolumn{2}{c}{$E^{\gamma}_{\mathrm{T}} > 40~\GeV$} \\
        & \multicolumn{2}{c}{$|\eta^\gamma|<2.37$, $\Delta R(\ell,\gamma)>0.7$} \\
        & \multicolumn{2}{c}{  $\epsilon_h^p < 0.5$}\\
\hline
  \end{tabular} 
  \caption{Definition of the fiducial region at the particle generator level. Here $p_{\mathrm{T}}^\nu$ is the transverse momentum of the neutrino from the $W$ boson decay; 		$N_\ell$ is the number of leptons per event; $\epsilon_h^p$ is the photon isolation fraction.
}
\label{tab:fiducialcut}
\end{table}

In order to test different hypothetical values of the parameter of interest, the test statistic based on the Profile Likelihood Ratio $\lambda(\sigma_{\mbox{\tiny{Fid}}}; \theta)$ is used, where $\theta$ is the vector of nuisance parameters,
defined in Ref.~\cite{PCLpaper}.
It is defined from the likelihood describing the probability density function of 
$m_{\mathrm{T}}^{\ell\nu\gamma}$ and $m^{\ell^{+}\ell^{-}\gamma}$ under a signal-plus-background hypothesis and is a function of the parameter of interest 
and of the nuisance parameters related to the systematic uncertainties. 
The statistical tests are then performed on the $m_{\mathrm{T}}^{\ell\nu\gamma}$ and $m^{\ell^{+}\ell^{-}\gamma}$ distributions combining the muon and electron channels
in each final state.

The probability of the background-only hypothesis, or local $p_{0}$-value, is obtained by using a frequentist approach.
The latter gives the probability that the background fluctuates to the observation or above.
If no hint for a new physics signal is found, the data is interpreted by using a modified frequentist approach ($CL_{s}$)~\cite{CLpaper} for setting limits. A fiducial cross section 
is claimed to be excluded at 95\% CL when $CL_{s}$ is less than 0.05.

\subsection{Systematic uncertainties}
\label{sec:syst}
Systematic uncertainties on the signal resonances are taken into account as nuisance parameters in the likelihood function used for the full signal and background model. Two different effects are evaluated for each source of systematic uncertainty, one for the signal event rate and one for the resolution of the signal. The rest of the parameters are kept fixed to their nominal value in the fit. 
Each systematic effect is investigated by propagating the corresponding uncertainty to the signal sample. These are computed separately for each of the simulated resonance mass points. The systematic uncertainties are summarized below for $m_{\tomega}=500\GeV$ in the \Zgamma\ channel and $m_{\techa}=600\GeV$ in the \Wgamma\ channel.

For the photon, systematic effects due to isolation, identification, energy resolution and
energy scale are considered. The systematic uncertainties due to the photon reconstruction and identification are dominant.  The total effect on the signal event rate is estimated to be $3.4\%$ in all of the channels, and contributes up to $3\GeV$, depending on the final state and channel, to the systematic uncertainty on the signal peak width.

The systematic effects due to the electron energy resolution and electron energy scale~\cite{egammascale} are treated as fully correlated with the photon energy scale and resolution in the final states containing electrons ($e^{+}e^{-}\gamma$ and $e\nu\gamma$). The effects of the muon energy scale and muon energy resolution, lepton identification and trigger efficiency are also investigated. The total effect of the lepton reconstruction and identification on the signal event rate reaches $1.5\%$ in the electron channels, and $1\%$ in the muon channels. The effect on the peak width due to lepton reconstruction and identification amounts to $0.3\GeV$ or less.

Systematic effects due to the jet energy scale and resolution and the calibration of the missing transverse energy only have an impact on the \Wgamma\ final state. These are found to cause uncertainties in the event rate of about $0.9\%$ and on the peak width of up to $0.2\GeV$. The jet energy scale and resolution effects on the \Zgamma\ final state are negligible.

Finally a systematic uncertainty on the resonance production rate due to the $2.8\%$ uncertainty on the integrated luminosity~\cite{ATLASlumi2013} is considered. 

The total systematic uncertainty on the event rate is found to be approximately 5\%, adding
all independent contributions in quadrature, for all the mass points in the two channels.
The systematic uncertainty on the peak width is found to be approximately 1.5 (0.5)~\GeV{} for the \Zgamma\ final state at $m_{\tomega}=500\GeV$ in the electron (muon) channel, and 2.8 (1.1)~\GeV{} at $m_{\techa}=600\GeV$ in the \Wgamma\ final state, in the electron (muon) channel.

Since the SM backgrounds are determined using a fit to the data, uncertainties on the detector resolution and physics object reconstruction or identification have a negligible effect on the total background yield extraction.
However, a systematic effect from the background modeling is investigated. The method considered consists of generating background--only ``Asimov'' datasets\footnote{An Asimov dataset is defined as being a single representative dataset replacing the ensemble of simulated datasets.}~\cite{PCLpaper}, and fitting them with the signal and background model in order to estimate the bias introduced on the fitted signal yield (``spurious signal''). 
In the $\Zgamma$ low mass region, it is found that the spurious signal reaches 25\% of the statistical uncertainty on the total background yield, in the rest of the regions of all $Z\gamma$ and $W\gamma$ final states the spurious signal is below 10\% and has a negligible impact. For these regions an additional systematic uncertainty is assigned, associating a nuisance parameter to the amount of spurious signal estimated in the above described procedure.

\section{Results}
\label{sec:results}
\begin{figure*}[h!]
\begin{center}
\subfigure[]{\includegraphics[width=8.5cm]{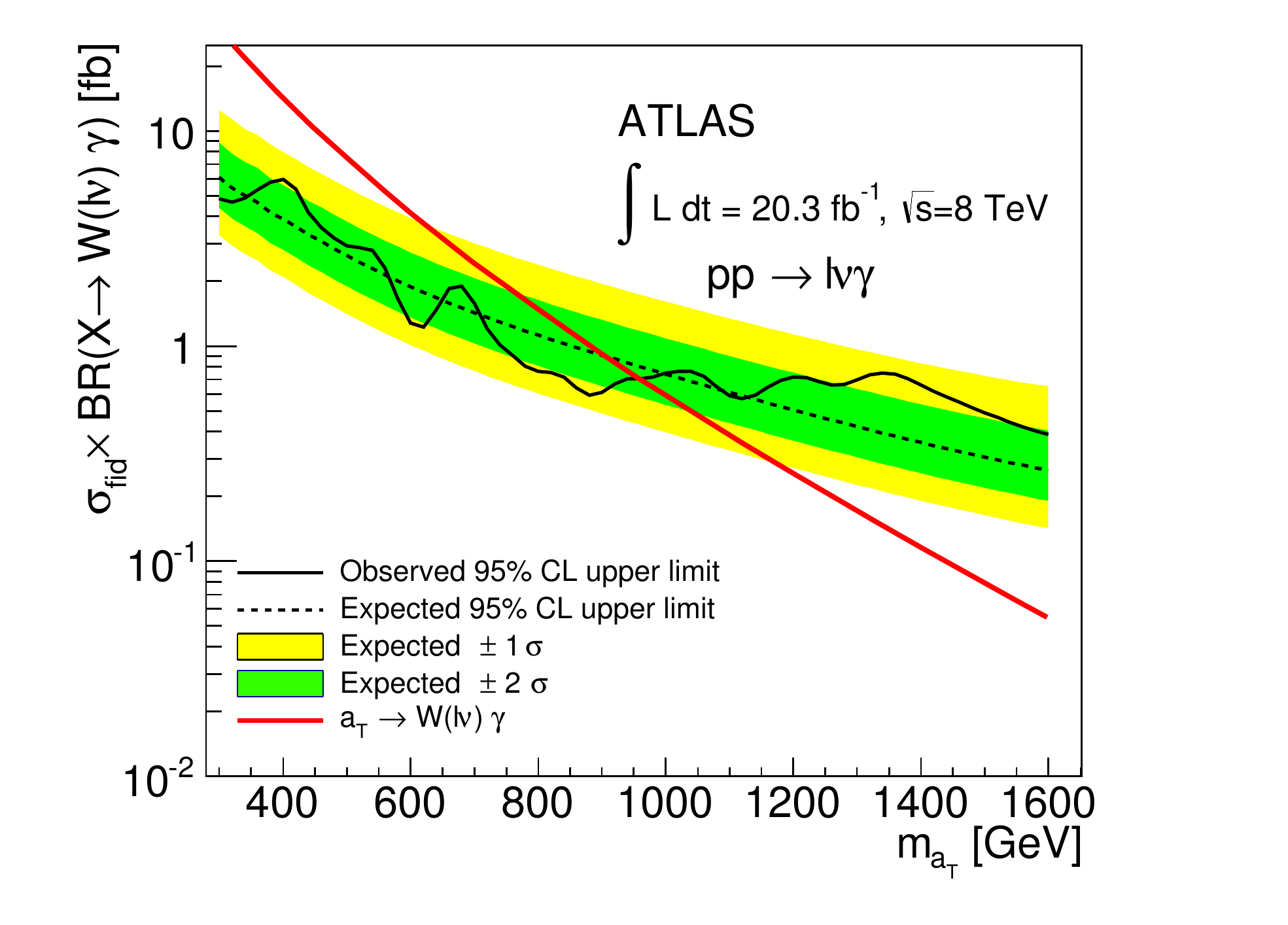}}
\subfigure[]{\includegraphics[width=8.5cm]{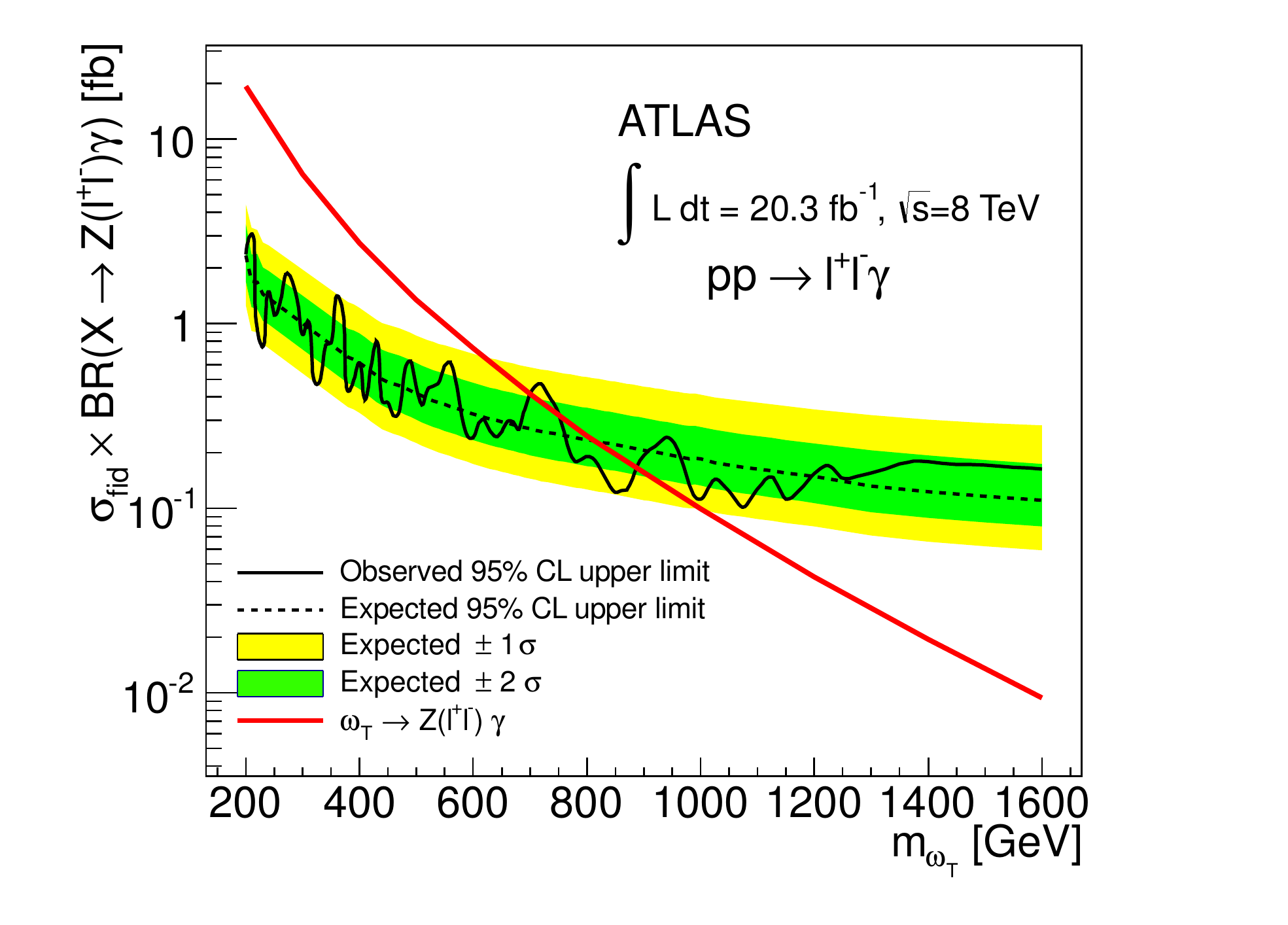}}
\subfigure[]{\includegraphics[width=8.5cm]{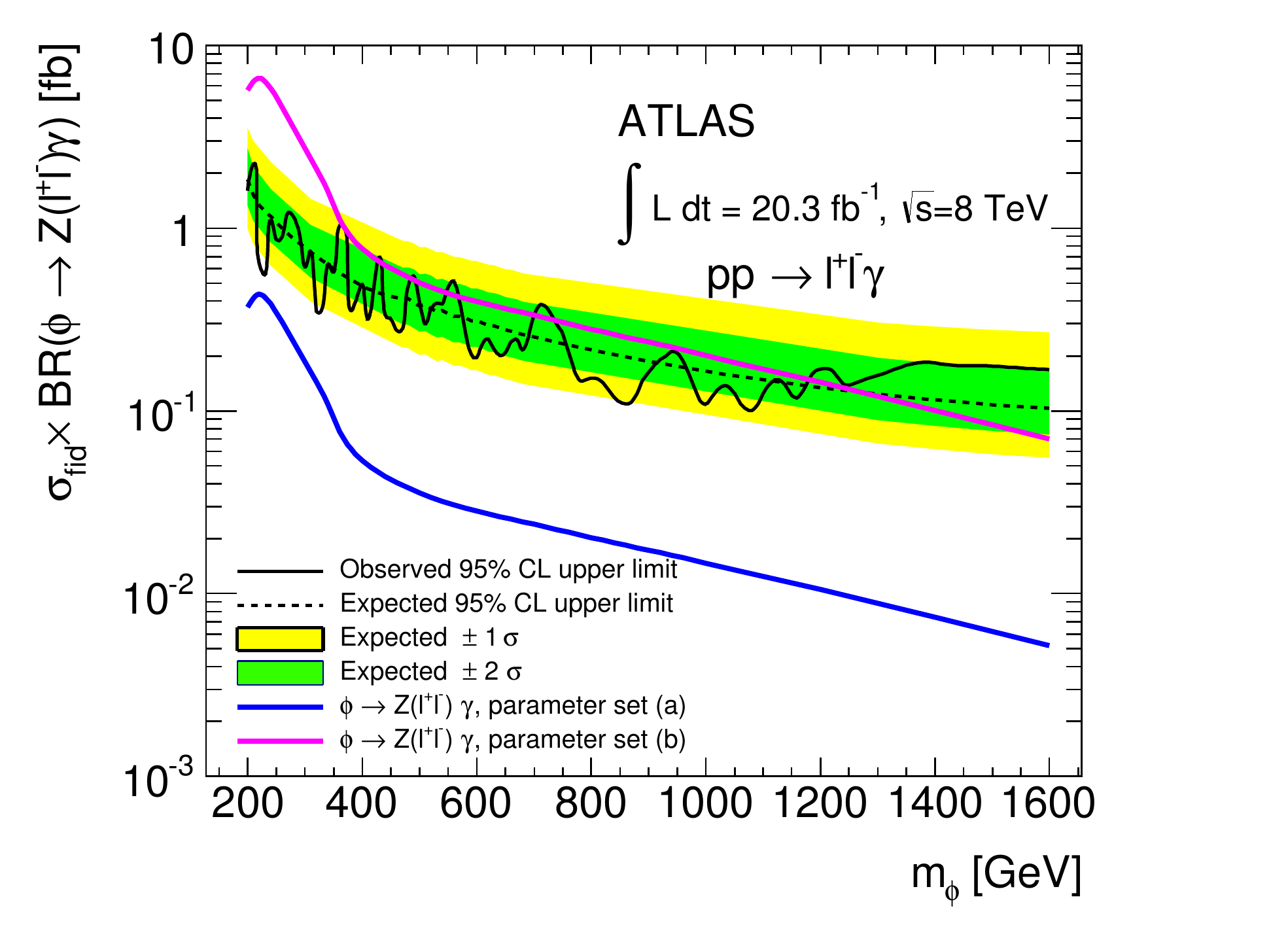}}
\end{center}
\caption{Observed limits in the fiducial region in the: $W\gamma$ final state as a function of $m_{\techa}$ (a), $Z\gamma$ final state as a function of $m_{\tomega}$ (b) and $Z\gamma$ final state as a function of $m_{\phi}$ (c) obtained using the full $8\TeV$ ATLAS dataset. The black line is the observed limit. The dashed black line is the expected limit, the yellow and green bands are the one and two $\sigma$ uncertainties on the expectation. The red lines in (a) and (b) are the theoretical cross sections for the LSTC benchmark models considered. The blue line and magenta line in (c) show the theoretical cross sections for the benchmark composite scalar model with the parameter set (a) and (b), respectively.  }
\label{fig:limitLSTC}
\end{figure*}

The two largest deviations from the background--only configuration are observed, in the spin-1 search, at $1350\GeV$ for the $\Wgamma$ final state, and at $350\GeV$ and $700\GeV$ for the $\Zgamma$ final state. Similarly, in the spin-0 search, the largest deviation is close to $350\GeV$. The $p_{0}$-values are estimated to be greater than $0.01$ in all cases, corresponding in all cases to a local significance below approximately $2.5\sigma$. These small excesses, also visible in the mass spectra in Figs.~\ref{fig:AllMassDistributionsW} and~\ref{fig:AllMassDistributionsZ}, are at the level expected due to statistical fluctuations.
No evidence of a new physics signal is found in the $W\gamma$ and $Z\gamma$ final states when analyzing the totality of the $8\TeV$ data collected by ATLAS in the mass ranges considered.

Therefore, $95\%$ CL limits are set on the production of LSTC technimesons and composite scalars decaying to $\Wgamma$ and $\Zgamma$ final states.
Figure~\ref{fig:limitLSTC} shows the observed and expected limits in the  fiducial region obtained for an LSTC $\techa \to \Wgamma$ (a), $\tomega \to \Zgamma$ (b),
and for a composite scalar $\phi \to \Zgamma$ (c). 
Using 20.3~fb$^{-1}$ of 8~\TeV{} data, the LSTC benchmark model is excluded at 95\% CL in the range $[275,960]\,\GeV$ for $W\gamma$, and $[200,700]\cup[750,890]\,\GeV$ 
for $Z\gamma$. In the spin-0 searches, no sensitivity is reached for the composite scalar model with the parameter set (a), whereas composite scalar models with the parameter set (b) are excluded at 95\% CL in most of the mass range $[200, 1180]\GeV$.

\section{Conclusions}
\label{sec:conclusions}

This paper describes the results of searches with the ATLAS detector
for LSTC technimesons and composite scalar particles decaying to $V\gamma$ final states.
The observed data prove to be consistent with the SM background expectations. 
Using the LSTC benchmark model, the production of $\techa{}$ is excluded up to $m_{\techa}=960~\GeV$ in the $W\gamma$ mode and the production of $\tomega{}$ is excluded in most of the mass range below $890~\GeV$ in the $\Zgamma$ channel. Using the benchmark composite scalar model, the production of $\phi$ is excluded in most of the mass range below $1180~\GeV$ for coupling scenario (b) described in text.
The limits on the LSTC benchmark are more stringent than the previous results obtained at ATLAS and documented in Ref.~\cite{WZgpaper}.


\section{Acknowledgements}

We thank CERN for the very successful operation of the LHC, as well as the
support staff from our institutions without whom ATLAS could not be
operated efficiently.

We acknowledge the support of ANPCyT, Argentina; YerPhI, Armenia; ARC,
Australia; BMWFW and FWF, Austria; ANAS, Azerbaijan; SSTC, Belarus; CNPq and FAPESP,
Brazil; NSERC, NRC and CFI, Canada; CERN; CONICYT, Chile; CAS, MOST and NSFC,
China; COLCIENCIAS, Colombia; MSMT CR, MPO CR and VSC CR, Czech Republic;
DNRF, DNSRC and Lundbeck Foundation, Denmark; EPLANET, ERC and NSRF, European Union;
IN2P3-CNRS, CEA-DSM/IRFU, France; GNSF, Georgia; BMBF, DFG, HGF, MPG and AvH
Foundation, Germany; GSRT and NSRF, Greece; ISF, MINERVA, GIF, I-CORE and Benoziyo Center,
Israel; INFN, Italy; MEXT and JSPS, Japan; CNRST, Morocco; FOM and NWO,
Netherlands; BRF and RCN, Norway; MNiSW and NCN, Poland; GRICES and FCT, Portugal; MNE/IFA, Romania; MES of Russia and ROSATOM, Russian Federation; JINR; MSTD,
Serbia; MSSR, Slovakia; ARRS and MIZ\v{S}, Slovenia; DST/NRF, South Africa;
MINECO, Spain; SRC and Wallenberg Foundation, Sweden; SER, SNSF and Cantons of
Bern and Geneva, Switzerland; NSC, Taiwan; TAEK, Turkey; STFC, the Royal
Society and Leverhulme Trust, United Kingdom; DOE and NSF, United States of
America.

The crucial computing support from all WLCG partners is acknowledged
gratefully, in particular from CERN and the ATLAS Tier-1 facilities at
TRIUMF (Canada), NDGF (Denmark, Norway, Sweden), CC-IN2P3 (France),
KIT/GridKA (Germany), INFN-CNAF (Italy), NL-T1 (Netherlands), PIC (Spain),
ASGC (Taiwan), RAL (UK) and BNL (USA) and in the Tier-2 facilities
worldwide.

\bibliographystyle{elsarticle-num}
\bibliography{plb.bbl}

\onecolumngrid
\clearpage

\begin{flushleft}
{\Large The ATLAS Collaboration}

\bigskip

G.~Aad$^{\rm 84}$,
B.~Abbott$^{\rm 112}$,
J.~Abdallah$^{\rm 152}$,
S.~Abdel~Khalek$^{\rm 116}$,
O.~Abdinov$^{\rm 11}$,
R.~Aben$^{\rm 106}$,
B.~Abi$^{\rm 113}$,
M.~Abolins$^{\rm 89}$,
O.S.~AbouZeid$^{\rm 159}$,
H.~Abramowicz$^{\rm 154}$
% ...
\bigskip
\\
% ...
$^{11}$ Institute of Physics, Azerbaijan Academy of Sciences, Baku, Azerbaijan\\
% ...
$^{84}$ CPPM, Aix-Marseille Universit{\'e} and CNRS/IN2P3, Marseille, France\\
% ...
$^{89}$ Department of Physics and Astronomy, Michigan State University, East Lansing MI, United States of America\\
% ...
$^{106}$ Nikhef National Institute for Subatomic Physics and University of Amsterdam, Amsterdam, Netherlands\\
% ...
$^{112}$ Homer L. Dodge Department of Physics and Astronomy, University of Oklahoma, Norman OK, United States of America\\
$^{113}$ Department of Physics, Oklahoma State University, Stillwater OK, United States of America\\
% ...
$^{116}$ LAL, Universit{\'e} Paris-Sud and CNRS/IN2P3, Orsay, France\\
% ...
$^{152}$ Institute of Physics, Academia Sinica, Taipei, Taiwan\\
% ...
$^{154}$ Raymond and Beverly Sackler School of Physics and Astronomy, Tel Aviv University, Tel Aviv, Israel\\
% ...
$^{159}$ Department of Physics, University of Toronto, Toronto ON, Canada\\
% ...
\end{flushleft}


\end{document}